\documentclass{article}
\usepackage{spconf,amsmath,amssymb,graphicx,booktabs,xcolor,hyperref}
\usepackage{orcidlink}

\hypersetup{hidelinks,pdfauthor={Shiqi Zhang, Tuomas Virtanen}}

\makeatletter
\renewcommand\section{\@startsection{section}{1}{\z@}%
  {-3.0ex \@plus -0.8ex \@minus -.2ex}%
  {1.8ex \@plus .2ex}%
  {\normalsize\bfseries}}
\makeatother
\DeclareMathOperator*{\argmax}{arg\,max}
\newcommand{\refrow}[1]{\textcolor{gray}{#1}}

\title{Geometric Consistency Based Label-Free Validation for Checkpoint Selection in Audio Active Learning}
\name{Shiqi Zhang\,\orcidlink{0009-0008-6012-8088}, Tuomas Virtanen\,\orcidlink{0000-0002-4604-9729}}
\address{Faculty of Information Technology and Communication Sciences\\
  Tampere University, Tampere, Finland}
\begin{document}
\ninept
\maketitle
\begin{abstract}
  Audio active learning (AL) trains a classifier at low annotation cost by labeling only an informative fraction of an unlabeled collection over successive rounds.
  Each round picks one of the checkpoints saved during training.
  Mainstream AL protocols do so with an assumed validation set, but in practice its labels also count toward the annotation cost, shrinking the advantage of an already small annotation budget.
  We propose geometric consistency, a label-free validation rule for checkpoint selection.
  Both classifiers, nearest-class-mean and \(k\)-nearest-neighbor, are built from the labeled set in a frozen pretrained embedding space.
  They then predict the unlabeled data to model its geometry.
  The rule scores each checkpoint by its agreement with these pseudo-labels and its cross-entropy on the labeled set.
  It works with any query strategy, has one hyperparameter, insensitive in practice, and adds only one batched forward pass over the unlabeled data per round.
  At matched label budgets across datasets, encoders, and query strategies, it attains the best mean rank in area under the labeling curve (AULC) among methods without extra labels and is on par with an out-of-budget validation set.
  To our knowledge, this is the first systematic study of label-free validation for checkpoint selection in audio AL.
\end{abstract}

\begin{figure*}[t]
  \centering
  \includegraphics[width=\textwidth]{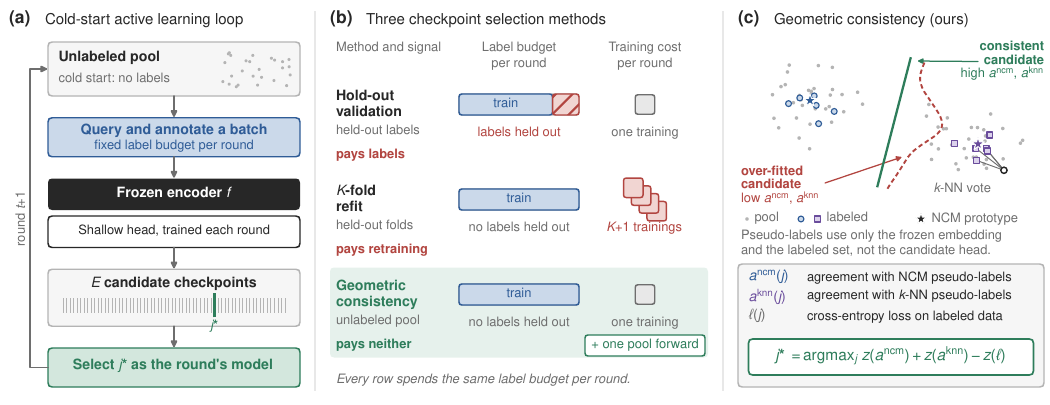}
  \caption{(a) Pool-based AL with a frozen encoder and a shallow head trained from scratch each round. A stopping method picks one of \(E\) candidate checkpoints as the model of the round. (b) Three checkpoint selection methods using the same per-round label budget. Hold-out validation leaves some labels for validation. \(K\)-fold refit does the validation with the same labels it then retrains on, which biases the selection, and requires extra retrainings. The proposed geometric consistency requires neither held-out labels nor retrainings and adds one batched pool forward. (c) Geometric consistency scores each checkpoint by the agreement of its predictions on the pool with the nearest-class-mean (NCM) and \(k\)-nearest-neighbor (\(k\)-NN) pseudo-labels and by its cross-entropy on the labeled set.}
  \label{fig:concept}
\end{figure*}

\section{Introduction}
\label{sec:intro}

Audio annotation is expensive, especially in specialized domains such as bioacoustics where it needs to be done by scarce domain experts.
Active learning (AL) contains this cost by labeling only an informative subset of a new audio collection. In each round, an acquisition strategy, the rule deciding which clips to annotate next, queries a batch of labels, and a classifier is retrained on all labels acquired so far.
On small labeled sets the model overfits easily, so performance depends on selecting which checkpoint during training.
Most AL studies make this choice with a ready-made validation set.
A real AL run starts without one, and validation labels have to be annotated like any other, so they consume the annotation budget.
Ali et al.~\cite{ali2014active} argued early on that model-selection labels should be counted as part of the budget, and that the performance reported under protocols that leave them outside is overestimated.
Throughout this paper we call a method \emph{deployable} when its checkpoint selection is done using only the labeled set and the unlabeled pool, the clips whose labels have not been queried, with no extra held-out labels.

Counting validation labeling as part of the budget leaves two established options, holding a validation split out of training or returning it to training through a refit, and each has a clear cost.
Holding a validation split out of the labeled data trades scarce supervision for a selection signal and shrinks the training set.
An alternative is to return the validation labels to training.
The checkpoint index is first selected using a temporary validation split, and the final model is then retrained on the full labeled set for the selected number of steps~\cite{prechelt1998early}.
A \(K\)-fold variant locates the checkpoint on the validation curves averaged over the folds and then performs the same refit~\cite{kohavi1995cv}.
Either refit trains the final model on the very labels that selected its checkpoint, so the validation estimate is no longer unbiased, and the \(K\)-fold variant further multiplies the per-round training cost.
Moreover, a validation split carved from a small labeled set covers the data poorly, so its distribution easily drifts from the true one.
Training for a fixed number of steps without any validation has been criticized as unreliable, since it leaves the classifier under-fit or over-fit~\cite{tan2024bemps}.
More fundamentally, the right step count varies with the dataset and the budget, and re-tuning it for each new collection requires exactly the labels the fixed schedule was meant to save.
A natural direction is therefore label-free validation, which replaces validation labels with a selection signal that costs no annotation.

Existing methods draw their signal from the training process and leave the unlabeled pool untouched.
General-purpose ones monitor training dynamics, such as gradient-weight alignment (GWA)~\cite{holzl2025gwa}, the number of training-set predictions that change between consecutive checkpoints in LabelWave~\cite{yuan2024labelwave}, and earlier gradient-evidence criteria~\cite{mahsereci2017early}. Besov early stopping (BEAST)~\cite{jukic2023smooth} instead analyzes how representations change across the layers of an end-to-end fine-tuned Transformer, an axis a frozen encoder does not provide.
Classic AL stopping criteria do use the unlabeled pool, through the cross-round stability of its predictions~\cite{bloodgood2009stopping}, yet they decide when to terminate the whole labeling process, not which checkpoint to select within a round.
Ali et al.~\cite{ali2014active} considered letting each candidate model pseudo-label the pool for model selection, but dismissed the scheme because a model scored on labels it produced yields a heavily biased error estimate.
The unlabeled pool is nonetheless a natural source of such a signal, since the acquisition strategy already utilizes its embeddings in every round and reusing them for selection costs no additional annotation.

Unsupervised validation in domain adaptation has shown that the feature geometry of unlabeled samples, such as their density and clustering, can rank checkpoints without labels~\cite{saito2021snd, yang2024transferscore}.
Inspired by this, we bring the principle to within-round checkpoint selection in AL and propose \emph{geometric consistency} (Geo-Con), which draws its signal from the pool instead of a validation set.
It scores each candidate checkpoint by the agreement of its predictions on the pool with two pseudo-labelers and by its cross-entropy on the labeled set.
The pseudo-labelers are a nearest-class-mean (NCM) and a \(k\)-nearest-neighbor (\(k\)-NN) classifier fitted on the labeled set in the frozen embedding space, so they carry both the embedding geometry and the label information of the task.
The rule selects the checkpoint maximizing the sum of the three standardized scores.
Both pseudo-labelers are independent of the candidate head under evaluation, which breaks the self-referential dependence that ruled the pool out before.

Under a unified budget protocol, where every method consumes exactly the same number of labels at every point of the labeling process, validation labels included, the proposed geometric consistency attains a mean rank of 1.08 in area under the labeling curve (AULC) among six deployable methods, ranks first in 45 of 48 settings, and outperforms holdout validation in every setting with a mean gain of 2.8\,pp AULC.
It performs on par with an out-of-budget validation set (mean gap 0.02\,pp), stays within 1.06\,pp of a test-informed oracle reference, and adds only one batched forward pass over the pool per round.
Our contributions are threefold.
(1) We propose geometric consistency, bringing pool geometry to within-round label-free validation in AL for the first time through head-independent NCM and \(k\)-NN pseudo-labelers, and ranking first on average among deployable methods.
(2) The rule is acquisition-agnostic and is evaluated with six representative strategies; to our knowledge this is the first systematic study of label-free validation for checkpoint selection in audio AL.
(3) We provide a unified-budget study over 4 datasets, 2 encoders, 6 strategies, and 10 seeds, where the rule performs on par with the out-of-budget validation reference.

\section{Problem Setup}
\label{sec:setup}

We consider pool-based AL that starts with an unlabeled set \(\mathcal{P}\) of audio clips.
We denote by \(\mathcal{L}_t\) and \(\mathcal{U}_t\) the labeled and the unlabeled subsets of \(\mathcal{P}\) after round \(t\), so \(\mathcal{L}_t\cup\mathcal{U}_t=\mathcal{P}\).
Below, we call \(\mathcal{U}_t\) the pool.
The process starts cold, with \(\mathcal{L}_0=\varnothing\) and \(\mathcal{U}_0=\mathcal{P}\), so no labeled data exists at the outset and a validation set can only be obtained by spending part of the annotation budget.
In each round \(t\), an acquisition strategy queries labels for \(n_q\) clips in \(\mathcal{U}_t\), so the cumulative budget, the total number of clips annotated up to round \(t\), is \(m_t=(t{+}1)\,n_q\).
A frozen encoder \(f\) maps every clip to an embedding, and a shallow classification head is trained from scratch on \(\mathcal{L}_t\) in every round.
The head parameters are saved at \(E\) intermediate training steps, giving candidate checkpoints \(\{\theta_j\}_{j=0}^{E-1}\), from which a stopping method selects one index \(j^{\star}\) as the model of the round.
This paper studies how to choose \(j^{\star}\) with no labels beyond \(\mathcal{L}_t\).
The selected model is also used for the next round's acquisition, so each stopping method produces its own labeling trajectory.

\section{Geometric Consistency}
\label{sec:method}

A frozen encoder pretrained on general-purpose audio maps clips with similar acoustic content to nearby embeddings, so clips of the same class tend to cluster and different classes tend to separate before any label of the task is seen.
The embeddings of the pool thus already reflect part of the class structure, and a good checkpoint should place its decision boundaries in accordance with this structure while fitting the labeled data.
An overfitted model memorizes individual samples, and its predictions on the pool drift away from the embedding geometry.
On the other hand, an underfitted model has not yet absorbed the label information.
In both cases, its agreement with pseudo-labelers that follow this geometry decreases.
Two complementary geometric pseudo-labelers provide the consistency measure, a global prototype view and a local neighborhood view, and the labeled-set cross-entropy provides the fit measure.
The prototype view can be built from very few labels per class and works from the earliest rounds, while the neighbor view assumes no unimodal clusters and complements it when within-class structure is rich (Fig.~\ref{fig:concept}c).

Let \(\mathbf{x}_i\) denote the embedding of clip \(i\), \(\cos(\cdot,\cdot)\) the cosine similarity, \(\mathcal{L}_t^c\) the set of labeled samples of class \(c\), and \(\mathcal{C}_t=\{c:|\mathcal{L}_t^c|>0\}\) the classes labeled by round \(t\).
The NCM labeler assigns each pool sample \(i\in\mathcal{U}_t\) the class whose prototype, the mean of the unit-normalized embeddings of the class, is most similar to it,
\begin{equation}
  y_i^{\mathrm{ncm}}=\argmax_{c\in\mathcal{C}_t}\cos(\mathbf{x}_i,\boldsymbol{\mu}_c),
  \quad
  \boldsymbol{\mu}_c=\frac{1}{|\mathcal{L}_t^c|}\sum_{r\in\mathcal{L}_t^c}\frac{\mathbf{x}_r}{\|\mathbf{x}_r\|}.
  \label{eq:ncm}
\end{equation}
The \(k\)-NN labeler votes over the \(k\) labeled samples \(\mathcal{N}_k(i)\subset\mathcal{L}_t\) most similar to \(i\), weighting each vote by its cosine similarity,
\begin{equation}
  y_i^{\mathrm{knn}}=\argmax_{c\in\mathcal{C}_t}\ \textstyle\sum_{r\in\mathcal{N}_k(i)}
  \big[\cos(\mathbf{x}_i,\mathbf{x}_r)\big]_{+}\,\mathbf{1}[y_r=c],
  \label{eq:knn}
\end{equation}
where \([u]_{+}=\max(u,0)\) clips negative similarities so that they cast no vote.
For each candidate checkpoint \(\theta_j\), we compute the agreement of its predicted labels on the pool with each pseudo-labeler, denoted \(a^{\mathrm{ncm}}(j)\) for the NCM labels and \(a^{\mathrm{knn}}(j)\) for the \(k\)-NN labels.
Agreement is the macro accuracy of these predictions against the pseudo-labels, averaged over the classes present in the pseudo-labels, so rare and common classes weigh equally, as in the evaluation metric.
The fit term \(\ell(j)\) is the mean cross-entropy loss on \(\mathcal{L}_t\) of the head at checkpoint \(j\).
The selected checkpoint maximizes the equally weighted sum of the three scores after z-standardization over the \(E\) checkpoints of the round,
\begin{equation}
  j^{\star}=\argmax_{j}\ z\big(a^{\mathrm{ncm}}(j)\big)+z\big(a^{\mathrm{knn}}(j)\big)-z\big(\ell(j)\big),
  \label{eq:rule}
\end{equation}
where \(z(\cdot)\) standardizes a score to zero mean and unit variance over the \(E\) checkpoints, which makes the three scores commensurate without per-dataset calibration; a score that is constant for all the checkpoints gets \mbox{\(z=0\)} and drops out of the ranking.

The only hyperparameter of the rule is the neighbor count \(k=10\) (ablated in Table~\ref{tab:ablation}).
The rule only uses the classifier outputs on the pool and the labeled set, so it is acquisition-agnostic and plugs into the standard AL loop unchanged.
Its overhead is one batched forward pass of all \(E\) checkpoints over the cached embeddings of the pool, modest for a two-layer head.

\section{Experimental Setup}
\label{sec:exp}

\begin{table}[t]
  \centering
  \caption{Aggregate results over 48 settings \(\times\) 10 seeds (mean\(\pm\)std across seeds, settings averaged within each seed). Rank averages within-setting AULC ranks of the six deployable methods. Gray rows are non-deployable references.}
  \label{tab:main}
  \small
  \setlength{\tabcolsep}{3.2pt}
  \begin{tabular}{lccc}
    \toprule
    Method             & Final acc (\%)            & AULC (\%)                 & Rank\(\downarrow\) \\
    \midrule
    GWA                & 74.53\(\pm\)0.24          & 69.94\(\pm\)0.30          & 4.17               \\
    LabelWave          & 75.13\(\pm\)0.34          & 69.82\(\pm\)0.34          & 3.69               \\
    holdout-v          & 74.99\(\pm\)0.24          & 69.19\(\pm\)0.31          & 5.25               \\
    holdout-v-refit    & 75.35\(\pm\)0.33          & 70.10\(\pm\)0.29          & 4.10               \\
    kfold-refit        & 76.03\(\pm\)0.27          & 70.99\(\pm\)0.20          & 2.71               \\
    Geo-Con (ours)     & \textbf{76.78\(\pm\)0.17} & \textbf{71.99\(\pm\)0.19} & \textbf{1.08}      \\
    \midrule
    \refrow{val-best}  & \refrow{76.76\(\pm\)0.18} & \refrow{72.01\(\pm\)0.27} & \refrow{---}       \\
    \refrow{test-best} & \refrow{77.94\(\pm\)0.15} & \refrow{73.05\(\pm\)0.17} & \refrow{---}       \\
    \bottomrule
  \end{tabular}
\end{table}

We run 4 datasets \(\times\) 2 encoders \(\times\) 6 strategies \(\times\) 8 stopping methods \(\times\) 10 seeds, 3840 runs in total.
Datasets are ESC-50 (50 classes)~\cite{piczak2015esc}, BSD10K (23)~\cite{anastasopoulou2024bsd}, DB3V (10)~\cite{jing2024db3v}, and UrbanSound8K (10)~\cite{salamon2014urbansound}; encoders are frozen CLAP (Transformer, contrastive pretraining, 512-d)~\cite{wu2023clap} and PANNs CNN14 (convolutional, supervised pretraining, 2048-d)~\cite{kong2020panns}.
Freezing the encoder mirrors the common audio AL setting~\cite{shishkin2024gdal, zhang2025hybrid} and holds the representation fixed across rounds, which isolates checkpoint selection, the focus of this study, from representation learning.
The four datasets span environmental sounds, heterogeneous Freesound content, cross-region bird vocalisations, and urban noise, with difficulty ranging from near-saturated pretrained features on ESC-50 to much weaker ones on BSD10K.
Acquisition covers uncertainty and diversity families: random, entropy~\cite{wang2014active}, margin~\cite{scheffer2001margin}, coreset~\cite{sener2018coreset}, MaxHerding~\cite{bae2024maxherding}, and BADGE~\cite{ash2020badge}.

For comparability, all experiments share one common default configuration.
Each run lasts 20 rounds with \(n_q=50\) (seed-paired random sampling at round~0 and the acquisition strategy in the other 19 rounds), up to 1000 labels.
The head is a two-layer MLP (hidden width 128, GELU) trained each round with Adam (learning rate \(10^{-2}\), batch size 32) for \(T_t=200\,\lceil m_t/32\rceil\) optimizer steps under a 20\% linear warmup followed by cosine decay.
\(T_t\) depends on the total budget \(m_t\) rather than on the split a method trains on, so the checkpoint step grid is aligned across methods at every budget point.
The \(E=100\) checkpoints are saved at steps \(s_j=\mathrm{round}\big((j{+}1)T_t/E\big)\), on optimizer steps rather than epochs, since a holdout model runs a non-integer number of epochs and only a step grid stays aligned across methods.

Baselines are GWA~\cite{holzl2025gwa} and LabelWave~\cite{yuan2024labelwave}, which use training dynamics only, and holdout-v, holdout-v-refit~\cite{prechelt1998early}, and kfold-refit~\cite{kohavi1995cv}, which need labels held out of training.
The holdout family spends 10 of the 50 labels per round (\(v{=}20\%\)) on validation and queries the other 40 with the acquisition strategy.
Following~\cite{ali2014active}, the validation clips are drawn uniformly at random from \(\mathcal{P}\) instead of by the acquisition strategy, so the validation pool is an unbiased sample of \(\mathcal{P}\).
The validation clips are chosen by one random permutation of \(\mathcal{P}\) before any label is acquired and are excluded from the query candidates.
holdout-v thus returns a model trained on 80\% of the labels.
holdout-v-refit reuses its selected step to retrain once on all labels.
kfold-refit selects the checkpoint on the validation curve averaged over a stratified 5-fold split and then retrains on all labels, so it sacrifices no labels but incurs \(K{+}1=6\) trainings per round.
All three, like the two references below, select the checkpoint with the highest macro accuracy on their held-out labels, with ties going to the earliest checkpoint.
Methods share no checkpoints or queries, only the cold-start batch of each seed.

\begin{table*}[t]
  \centering
  \caption{Test macro AULC (\%) per dataset \(\times\) encoder block (mean\(\pm\)std across seeds, strategies averaged within each seed). Best deployable method per column in bold; gray rows are non-deployable references.}
  \label{tab:blocks}
  \small
  \setlength{\tabcolsep}{4.5pt}
  \begin{tabular}{lcccccccc}
    \toprule
                       & \multicolumn{2}{c}{ESC-50} & \multicolumn{2}{c}{BSD10K} & \multicolumn{2}{c}{DB3V}  & \multicolumn{2}{c}{UrbanSound8K}                                                                                                                 \\
    \cmidrule(lr){2-3}\cmidrule(lr){4-5}\cmidrule(lr){6-7}\cmidrule(lr){8-9}
    Method             & CLAP                       & PANNs                      & CLAP                      & PANNs                            & CLAP                      & PANNs                     & CLAP                      & PANNs                     \\
    \midrule
    GWA                & 91.66\(\pm\)0.41           & 84.10\(\pm\)0.43           & 54.28\(\pm\)0.44          & 47.83\(\pm\)1.08                 & 60.87\(\pm\)1.14          & 56.02\(\pm\)1.06          & 83.56\(\pm\)0.60          & 81.20\(\pm\)0.61          \\
    LabelWave          & 91.91\(\pm\)0.63           & 84.83\(\pm\)0.40           & 53.74\(\pm\)0.73          & 48.46\(\pm\)0.84                 & 56.98\(\pm\)1.10          & 57.51\(\pm\)0.72          & 83.22\(\pm\)1.06          & 81.89\(\pm\)0.66          \\
    holdout-v          & 90.25\(\pm\)0.69           & 82.97\(\pm\)0.81           & 51.04\(\pm\)0.78          & 46.83\(\pm\)0.66                 & 60.62\(\pm\)1.17          & 57.11\(\pm\)1.01          & 83.38\(\pm\)0.90          & 81.31\(\pm\)0.40          \\
    holdout-v-refit    & 91.38\(\pm\)0.67           & 83.64\(\pm\)0.50           & 53.31\(\pm\)0.78          & 47.96\(\pm\)0.70                 & 62.33\(\pm\)0.76          & 57.01\(\pm\)0.92          & 83.81\(\pm\)1.01          & 81.35\(\pm\)0.50          \\
    kfold-refit        & 92.28\(\pm\)0.40           & 84.64\(\pm\)0.32           & 53.84\(\pm\)0.57          & 48.23\(\pm\)0.85                 & 63.80\(\pm\)0.34          & 59.03\(\pm\)0.81          & 84.16\(\pm\)0.67          & 81.89\(\pm\)0.42          \\
    Geo-Con (ours)     & \textbf{92.51\(\pm\)0.37}  & \textbf{85.00\(\pm\)0.40}  & \textbf{54.65\(\pm\)0.59} & \textbf{49.85\(\pm\)0.83}        & \textbf{65.41\(\pm\)0.40} & \textbf{60.89\(\pm\)0.84} & \textbf{84.55\(\pm\)0.61} & \textbf{83.07\(\pm\)0.60} \\
    \midrule
    \refrow{val-best}  & \refrow{92.45\(\pm\)0.38}  & \refrow{85.24\(\pm\)0.40}  & \refrow{54.53\(\pm\)0.57} & \refrow{49.71\(\pm\)0.81}        & \refrow{65.18\(\pm\)0.44} & \refrow{60.93\(\pm\)0.93} & \refrow{85.09\(\pm\)0.60} & \refrow{82.94\(\pm\)0.56} \\
    \refrow{test-best} & \refrow{93.03\(\pm\)0.38}  & \refrow{86.25\(\pm\)0.39}  & \refrow{55.48\(\pm\)0.65} & \refrow{50.49\(\pm\)0.77}        & \refrow{66.15\(\pm\)0.52} & \refrow{62.01\(\pm\)0.52} & \refrow{86.53\(\pm\)0.52} & \refrow{84.51\(\pm\)0.46} \\
    \bottomrule
  \end{tabular}
\end{table*}

Evaluation uses test macro accuracy at each budget point and the macro area under the labeling curve (AULC), a normalized trapezoidal integral of macro accuracy against cumulative labels.
All eight stopping methods consume exactly the same number of labels at every budget point, validation labels included, following the accounting principle of~\cite{ali2014active}; we call this protocol the unified budget.
Under this accounting, a stopping method pays for its selection signal in one of three forms: labels held out from the budget, extra retrainings of the head, or, as in our method, one batched forward pass over the unlabeled pool (Fig.~\ref{fig:concept}b).
Two references cross the deployable boundary: \emph{val-best} selects with the i.i.d.\ validation split of each dataset (400 to 1147 labeled clips, on the order of the whole budget), and \emph{test-best} greedily reads test labels each round as an oracle reference.
Aggregates average the 48 settings within each seed and report mean\(\pm\)std across seeds, and within-setting ranks compare the 10-seed mean AULC of the six deployable methods.

\section{Results}
\label{sec:results}

\textbf{Main results.}
Table~\ref{tab:main} aggregates all 48 settings.
Geo-Con reaches 71.99\% AULC, ahead of the runner-up kfold-refit by 1.01\,pp (better in 47 of 48 settings) and ahead of holdout-v by 2.80\,pp (better in all 48).
Its final macro accuracy of 76.78\% is the best among deployable methods as well, and its mean rank of 1.08 comes from ranking first in 45 of 48 settings.
Against the references it is on par with val-best (72.01\%, mean gap 0.02\,pp, ahead in 27 of 48 settings) and is 1.06\,pp below the test-informed test-best reference, so the pool signal picks checkpoints almost as well as an i.i.d.\ validation set.
GWA and LabelWave, the two training-dynamics baselines, trail by roughly 2\,pp AULC, so using the pool brings gains beyond monitoring the training process.
holdout-v and holdout-v-refit share the same selection step and differ only in whether the validation labels return to training, so their 0.91\,pp AULC gap measures the cost of the training set that validation shrinks.
On near-saturated settings such as ESC-50 with CLAP, final-accuracy differences compress while AULC still separates convergence speed.

\textbf{Across datasets and encoders.}
Table~\ref{tab:blocks} splits the comparison into eight dataset-encoder blocks; Geo-Con is the best deployable method in every block, and its margin tends to grow with task difficulty.
The AULC gain over holdout-v is largest on DB3V (+4.28\,pp), followed by BSD10K (+3.31), ESC-50 (+2.15), and UrbanSound8K (+1.46).
Its per-block gap to val-best stays within 0.25\,pp in seven of eight blocks, the largest deficit being 0.54\,pp on UrbanSound8K with CLAP.
The only 3 non-first settings of 48 (at most 0.96\,pp behind) sit at this low-gain edge, two on UrbanSound8K with CLAP.

\begin{table}[t]
  \centering
  \caption{Mean AULC rank of the six deployable methods under each acquisition strategy (8 settings per strategy).}
  \label{tab:strategies}
  \small
  \setlength{\tabcolsep}{3.5pt}
  \begin{tabular}{lcccccc}
    \toprule
    Method          & rand.         & entr.         & marg.         & core.         & mherd.        & badge         \\
    \midrule
    GWA             & 4.00          & 4.38          & 4.38          & 4.00          & 3.75          & 4.50          \\
    LabelWave       & 3.88          & 4.38          & 3.50          & 3.50          & 3.63          & 3.25          \\
    holdout-v       & 5.38          & 5.25          & 5.38          & 5.63          & 4.75          & 5.13          \\
    holdout-v-refit & 4.13          & 3.50          & 4.13          & 3.38          & 5.13          & 4.38          \\
    kfold-refit     & 2.63          & 2.50          & 2.50          & 3.25          & 2.63          & 2.75          \\
    Geo-Con (ours)  & \textbf{1.00} & \textbf{1.00} & \textbf{1.13} & \textbf{1.25} & \textbf{1.13} & \textbf{1.00} \\
    \bottomrule
  \end{tabular}
\end{table}

\textbf{Compatibility with acquisition strategies.}
By construction the rule is agnostic to the acquisition strategy, and Table~\ref{tab:strategies} confirms this empirically. Its mean rank stays at or below 1.25 under all six strategies, plain random sampling included, so the advantage does not hinge on any particular query strategy.
kfold-refit is the runner-up under every strategy, at six trainings per round.
The holdout family is last under every strategy, and holdout-v under five of the six, so validation labels cost more than any strategy recovers.

\begin{table}[t]
  \centering
  \caption{Ablations over the full matrix (48 settings \(\times\) 10 seeds, mean\(\pm\)std across seeds). Each variant replaces Geo-Con in the ranking against the five deployable baselines. Top, component ablation, best per column in bold; bottom, sensitivity to \(k\) (default \(k{=}10\) is the full rule).}
  \label{tab:ablation}
  \small
  \setlength{\tabcolsep}{3pt}
  \begin{tabular}{lccc}
    \toprule
    Variant                               & Final acc (\%)            & AULC (\%)                 & Rank\(\downarrow\) \\
    \midrule
    full rule, Eq.~(\ref{eq:rule})        & \textbf{76.78\(\pm\)0.17} & \textbf{71.99\(\pm\)0.19} & \textbf{1.08}      \\
    no fit term \(\ell\)                  & 75.09\(\pm\)0.28          & 71.22\(\pm\)0.21          & 1.77               \\
    no \(k\)-NN term \(a^{\mathrm{knn}}\) & 76.64\(\pm\)0.25          & 71.93\(\pm\)0.20          & 1.10               \\
    no NCM term \(a^{\mathrm{ncm}}\)      & 76.74\(\pm\)0.17          & 71.82\(\pm\)0.19          & 1.31               \\
    \(a^{\mathrm{ncm}}\) only             & 74.44\(\pm\)0.32          & 70.77\(\pm\)0.21          & 2.27               \\
    \(a^{\mathrm{knn}}\) only             & 76.26\(\pm\)0.16          & 71.50\(\pm\)0.19          & 1.77               \\
    \(\ell\) only                         & 76.56\(\pm\)0.17          & 71.74\(\pm\)0.22          & 1.48               \\
    \midrule
    \(k{=}5\)                             & 76.75\(\pm\)0.16          & 71.96\(\pm\)0.22          & 1.17               \\
    \(k{=}10\) (default)                  & \textbf{76.78\(\pm\)0.17} & \textbf{71.99\(\pm\)0.19} & \textbf{1.08}      \\
    \(k{=}20\)                            & \textbf{76.78\(\pm\)0.17} & 71.92\(\pm\)0.24          & 1.23               \\
    \bottomrule
  \end{tabular}
\end{table}

\textbf{Ablations.}
Table~\ref{tab:ablation} ablates the three components and the hyperparameter \(k\) on the full matrix.
Removing any component costs performance, though by different amounts.
Dropping the fit term \(\ell\) hurts most (\(-0.77\)\,pp AULC, rank 1.77), dropping the NCM term \(a^{\mathrm{ncm}}\) is second (\(-0.17\), rank 1.31), and dropping the \(k\)-NN term \(a^{\mathrm{knn}}\) is smallest (\(-0.07\), rank 1.10).
On top of the fit term alone, the two geometric terms add 0.25\,pp AULC and improve the rank from 1.48 to 1.08, with the NCM term carrying most of the gain.
All seven variants still hold the best mean rank against all baselines, indicating that pool geometry itself is an effective signal.
The ablation also reveals an interesting phenomenon.
The fit term alone, the loss on the labeled set, outperforms every baseline and also the \(a^{\mathrm{ncm}}\)-only and \(a^{\mathrm{knn}}\)-only variants.
We attribute this to the shallow MLP head, whose parameter freedom and feature-space capacity are limited compared with a deep end-to-end network, so even when it overfits, the symptoms are milder.
For training efficiency, this preliminary study uses the fixed-embedding plus MLP protocol common in audio AL, and an end-to-end investigation is left for future work.
Varying \(k\) over a fourfold range changes AULC by at most 0.07\,pp, and every value keeps the first deployable rank and stays on par with val-best.

\section{Conclusion}
\label{sec:conclusion}

Validation annotation budget is a hidden cost of mainstream AL protocols, while the proposed geometric consistency rule takes its selection signal from the unlabeled pool's geometry at no cost in labels or retraining.
It ranks first among deployable methods across 3840 runs and performs on par with an out-of-budget validation set.
These results suggest that the unlabeled pool can be a source of reliable model-selection signals.
Extending the rule to end-to-end fine-tuning, where the geometry moves with every checkpoint, and to multi-label tagging is a potential direction for future work.
We recommend that future audio AL evaluations count validation labels toward the annotation budget.

\clearpage


\section{Compliance with Ethical Standards}
\label{sec:ethics}
This study is limited to computational analysis of the existing public ESC-50~\cite{piczak2015esc}, BSD10K~\cite{anastasopoulou2024bsd}, DB3V~\cite{jing2024db3v}, and UrbanSound8K~\cite{salamon2014urbansound} datasets, for which no ethical approval was required.
No new experiments involving human or animal subjects were conducted.

\section{Acknowledgements}
\label{sec:acknowledgements}
This work received funding from the European Union's Horizon Europe programme under the Marie Sk{\l}odowska-Curie grant agreement No.~101116715 (BioacousticAI).

\bibliographystyle{IEEEbib}
\bibliography{refs}

\end{document}